\documentclass[aps,twocolumn,prd,notitlepage,amssymb,amsmath,floatfix,nofootinbib,superscriptaddress]{revtex4-1}

\usepackage{epsfig}
\usepackage{bm}
\usepackage{amssymb}
\usepackage{amsmath}
\usepackage{color}
\usepackage{subfigure}
\usepackage[colorlinks,linkcolor=blue,anchorcolor=black,citecolor=blue]{hyperref}

\allowdisplaybreaks[4]

\begin{document}

\title{Exploring non-perturbative Sudakov factor via $Z^0$-boson production in $pp$ collisions}

\author{Shu-yi Wei}   
\email{swei@ectstar.eu}
\affiliation{European Centre for Theoretical Studies in Nuclear Physics and Related Areas (ECT*) and Fondazione Bruno Kessler, Strada delle Tabarelle 286, I-38123 Villazzano (TN), Italy}

\begin{abstract}
$Z^0$-boson production at low transverse momentum offers an unique opportunity to explore the non-perturbative Sudakov factor. In this paper, we employ three parameterizations of the non-perturbative Sudakov factor to calculate the $\phi^*$-distribution and compare our results with the ultra precise experimental data. We extract the free parameters in each parameterization with a $\chi^2$ analysis. The parton-momentum-fraction dependence of these free parameters is also studied by comparing the values extracted at different collision energies and different rapidity ranges.
\end{abstract}

\maketitle

\section{Introduction}

In the collinear factorization framework, multiple soft gluon radiation, namely parton shower, contributes to large logarithms that destroy the predictive power of perturbative expension. A systemic resummation of these logarithms at all orders in QCD, i.e. Sudakov resummation, is thus vital in calculating observables that are sensitive to the parton shower effect. 

The Sudakov resummation formalism was first established for dihadron production in $e^+e^-$ annihilations and Drell-Yan process \cite{Collins:1981uk,Collins:1981va,Collins:1984kg}. Recently this formalism was also extended to more complicated processes, such as semi-inclusive deep inelastic scatterings (SIDIS) and dijet production in $pp$ collisions \cite{Ji:2004wu,Ji:2004xq,Sun:2014gfa,Sun:2015doa}, where soft gluons can be radiated from initial state and final state partons. 

Besides these charming therotical progresses, the applications of Sudakov resummation in phenomenology have also obtained great success. In Refs.~\cite{Zhang:2002yz,Kang:2012am,Kang:2016ron}, it has been employed to study the cold nuclear effect in $pA$ collisions. In Refs.~\cite{Mueller:2016gko,Chen:2016vem,Chen:2016cof,Chen:2018fqu}, it has been used to extract the jet transport coefficient in the quark-gluon plasma in the relativistic heavy-ion collisions. Recently, Sudakov resummation has also been incorporated into the dilute-dense factorization to probe the gluon saturation physics \cite{Mueller:2013wwa, Sun:2014gfa, Mueller:2016xoc,Zheng:2014vka,Stasto:2018rci,Marquet:2019ltn}. In those studies, Sudakov resummation establishes baselines for observables in $pp$ collisions. On top of that, relevant effects in $pA$ or $AA$ collisions can then be investigated. The uncertainties in the Sudakov resummation can also propagate into the final results and may have some impact on the conclusion. Therefore it should be dealt with carefully. 

The Sudakov factor, which is the most essential ingredient of the Sudakov resummation, consists of a perturbative part and a non-perturbative part. The perturbative part resums the large logarithms which can be calculated up to leading logarithm (LL) accuracy or next-to-leading logarithm (NLL) accuracy without any ambiguities. The non-perturbative part can only be determined by the experimental measurements and therefore brings vagueness in the theoretical evaluation. Thanks to the universality \cite{Collins:2004nx,Konychev:2005iy}, such uncertainty can be significantly reduced through a global analysis. 

On the other hand, in connection with the transverse-momentum-dependent factorization ($k_t$-dependent fractorization) framework \cite{Ji:2004wu,Ji:2004xq}, the non-perturbative Sudakov factor can be considered as the Fourier transform of the transverse momentum distribution at the initial scale, while the perturbative Sudakov factor evolves this distribution to the resummation scale. The study of non-perturbative Sudakov factor can also help us to draw the three-dimension picture of hadron. Recently, the $k_t$--dependent parton distribution functions have also been studied in Refs.~\cite{Martinez:2019mwt,Hautmann:2020cyp} from $Z^0$-boson production using the parton branching Monte Carlo method \cite{Hautmann:2017fcj}.

Several parameterizations \cite{Davies:1984sp,Landry:2002ix,Su:2014wpa,Ladinsky:1993zn} have been proposed for the non-perturbative Sudakov factor, where the values of the corresponding parameters are extracted from a global analysis to the expermental measurements of SIDIS \cite{Airapetian:2012ki,Adolph:2013stb} and Drell-Yan process \cite{Ito:1980ev,Antreasyan:1981uv,Moreno:1990sf,Affolder:1999jh,Abbott:1999wk,Abazov:2007ac,Aaltonen:2012fi,Aad:2011gj}. The transverse momentum spectrum ($q_T$-spectrum) of the $Z^0$-boson production in $pp$ collisions \cite{Affolder:1999jh,Abbott:1999wk,Abazov:2007ac,Aaltonen:2012fi,Aad:2011gj} is usually included in the analysis as well. However, it is still far from being the decisive factor due to the following two reasons. First, the uncertainty in the experimental measurement on the the $q_T$-spectrum of $Z^0$-boson is normally very large. Second, the Sudakov factor is overwhelmed by the perturbative logarithms which makes the $q_T$-spectrum not very sensitive to the non-perturbative physics. 

A new observable, $\phi^*$, has been proposed in \cite{Banfi:2010cf,Banfi:2012du} for $Z^0$-boson production to optimize the previous measurements on the $q_T$-spectrum recently. The uncertainty in the $\phi^*$ distribution mainly arises from the statistics instead of the detector's energy resolution which is the dominate contribution to the error bars in the $q_T$-distribution. It has been shown in Refs.~\cite{Aad:2012wfa,Abazov:2010mk} the measurements on the $\phi^*$-distribution are extremely precise, which makes it possible to explore non-perturbative physics from $Z^0$-boson production. In Ref.~\cite{Guzzi:2013aja}, the $\phi^*$-distribution has been used to determine the Gaussian width of non-perturbative Sudakov factor at the Tevatron energy.

In most of the parameterizations \cite{Davies:1984sp,Landry:2002ix,Su:2014wpa,Ladinsky:1993zn} of non-perturbative Sudakov factor in the market, the free parameters are usually assumed to be functions of the partonic center-of-mass energy, $Q$, and a combination of parton momentum fractions, $x_1x_2$. Unlike the case for low energy Drell-Yan process, $Q=M_Z$ is fixed for $Z^0$-boson production at low-$q_T$ since the $Z^0$-pole is quite steep. Thus, we have removed one variable from the equation and can simply study the parton momentum fraction dependence. This unique feature has granted $Z^0$-boson production an irreplaceable advantage in the study of non-perturbative Sudakov factor.

This paper is organised as follows. In Sec.~II, we demonstrate the formalism to calculate $\phi^*$-distribution of $Z^0$-boson production in $pp$ collisions. In Sec.~III, we show the parameterizations of non-perturbative Sudakov factor. In Sec.~IV, we present the numerical results. A summary is given in Sec.~V.

\section{Formalism}

The Sudakov resummation formalism of the differential $Z^0$-boson production cross section at low-$q_T$ in $pp$ collisions has been established by Collins, Soper and Sterman \cite{Collins:1984kg} decades ago and a lot of phenomenological researches have been carried out since then \cite{Davies:1984sp,Arnold:1990yk,Ladinsky:1993zn,Ellis:1997sc,Qiu:2000ga,Qiu:2000hf,Landry:2002ix,Bozzi:2008bb,Bozzi:2010xn,Hautmann:2012sh,Sun:2013hua,Su:2014wpa,Peng:2014hta,Catani:2015vma,Bizon:2018foh,Blanco:2019qbm,Marquet:2019ltn}. 

In the $q_T$-measurements, the uncertainty mainly arises from the energy resolution of the detectors. To improve the precision, $\phi^*$ is proposed as a novel observable to be measured in experiments. It can be easily measured from the polar and azimuthal angles of the final state dilepton. We only need to make sure the invariant mass of the final state dilepton pair is around the $Z^0$-pole, while the exact value of $q_T$ is not needed. Therefore the precision is mainly limited by the statistics. 

At low-$q_T$ limit, $\phi^* \simeq |q_T^y|/M_Z$ measures one component of the transverse momentum. It is quite straightforward to derive the $\phi^*$-distribution that is given by \cite{Marquet:2019ltn}
\begin{align}
\frac{d\sigma}{dy d\phi^*} = & 2M_Z \int_{-\infty}^{\infty} \frac{db_\perp}{2\pi} \frac{4\pi^2\alpha}{3s} \cos(|b_\perp| M_Z \phi^*) \nonumber \\
& \sum_{a,b,i} Q^2_{i\bar i} e^{-S_{\rm sud} (M_Z, |b_\perp|)} 
\nonumber \\
& \int \frac{d\xi_1}{\xi_1} C_{i(\bar i),a}(x_1/\xi_1) f_{a/A}(\xi_1,\mu_b) 
\nonumber \\
& \int \frac{d\xi_2}{\xi_2} C_{\bar i (i),b}(x_2/\xi_2) f_{b/B}(\xi_2,\mu_b),
\label{eq:cs-phistar}
\end{align}
where, $f_{a/A}(\xi_1,\mu_b)$ is the collinear parton distribution function (PDF) \cite{Dulat:2015mca} with $x_{1,2}=M_Z e^{\pm y}/\sqrt{s}$ the parton momentum fraction and $C_{i,a}(x_1/\xi_1) = \sum_{n=0} C_{i,a}^{(n)} (z=x_1/\xi_1) (\frac{\alpha_s (\mu_b)}{\pi})^{n}$ the coefficient function which can be calculated perturbatively \cite{Collins:1984kg, Davies:1984sp, Qiu:2000hf}, $Q^2_{i\bar i} = \frac{(1-4|e_i|\sin^2 \theta_W)^2 +1}{16 \sin^2\theta_W \cos^2 \theta_W}$ is the coupling constant, where $i\bar i$ labels a pair of quark and anti-quark with the same flavor, and $\mu_b = 2e^{-\gamma_E}/|b_\perp|$ with 
$\gamma_E$ 
the Euler constant. At the $\alpha_s^1$ order, the coefficient functions are given by 
\begin{align}
& C_{i,j}^{(0)} (z) = \delta_{ij} \delta (z-1), 
\quad\quad
C_{i,g}^{(0)} (z) = 0, \\
& C_{i,j}^{(1)} (z) = \delta_{ij} \left( \frac{2}{3} (1-z) + \delta (z-1) (\frac{\pi^2}{3} - \frac{8}{3}) \right), \\
& C_{i,g}^{(1)} (z) = \frac{1}{2} z (1-z). 
\end{align}

Sudakov factor is an integral over the scale from $\mu_b$ to $M_Z$. In phenomenology, we usually need to introduce the $b_*$-prescription to prevent this integral from going to the non-perturbative region. By setting $b_* = |b_\perp|/\sqrt{1+b_\perp^2/b_{\rm max}^2}$ and $\mu_b = 2e^{-\gamma_E}/b_*>2e^{-\gamma_E}/b_{\rm max}$, with $b_{\rm max}=0.5$ GeV$^{-1}$ the infrared cutoff, the non-perturbative sector is discarded from the integral. It is a common practise to add a non-perturbative Sudakov factor to take into account the missing effect. Therefore the Sudakov factor in Eq.~(\ref{eq:cs-phistar}) becomes
\begin{align}
S_{\rm sud} (M_Z, |b_\perp|) = S_{\rm pert} (M_Z, |b_\perp|)  + S_{\rm np}.
\end{align}
The perturbative Sudakov factor at the NLL accuracy is given by
\begin{align}
S_{\rm pert} = \int_{\mu_b^2}^{M_Z^2} \frac{d\mu^2}{\mu^2} 2\left[ \ln\frac{M_Z^2}{\mu^2} (A_1 + A_2) + B_1 \right],
\end{align}
with $A_1= C_F \frac{\alpha_s(\mu)}{2\pi}$, $A_2 = C_F \frac{\alpha_s^2(\mu)}{4\pi^2} [(\frac{67}{18}-\frac{\pi^2}{6})C_A - \frac{6}{9}N_f]$ and $B_1 = -\frac{3}{2} C_F \frac{\alpha_s(\mu)}{2\pi}$, where $C_F =4/3$ and $C_A=3$ are the color factors. The non-perturbative Sudakov factor, $S_{\rm np}$, is still yet to be extracted.

$b_*$-prescriotion is not the only method which can be employed in phenomenology. Other approaches to bridge the large logarithms at small-$b_\perp$ and the non-perturbative physics at large-$b_\perp$ also exist. See the discussion in Ref.~\cite{Qiu:2000hf} for an example. In this work, we stay in the context of $b_*$-prescription.

\section{Parameterization of non-perturbative Sudakov factor}

Several parameterizations of the non-perturbative Sudakov factor are available by far. We summarize the general forms in the following. 
\begin{itemize}

\item The Davies-Webber-Stirling (DWS) parameterization \cite{Davies:1984sp}:
\begin{align}
S_{\rm np}^{\rm DWS} = (g_1 + g_2 \ln \frac{Q}{2Q_0}) b_\perp^2.
\end{align}

\item The Brock-Landry-Nadolsky-Yuan (BLNY) parameterization \cite{Landry:2002ix}:
\begin{align}
S_{\rm np}^{\rm BLNY} = (g_1 + g_2 \ln \frac{Q}{2Q_0} + g_3 \ln (100x_1x_2)) b_\perp^2,
\end{align}

\item The Sun-Isaacson-Yuan-Yuan (SIYY) parameterization \cite{Su:2014wpa}:
\begin{align}
S_{\rm np} = g_1 b_\perp^2 + g_2 \ln\frac{Q}{Q_0} \ln\frac{|b_\perp|}{b_*}.
\end{align}

\item The Ladinsky-Yuan (LY) parameterization \cite{Ladinsky:1993zn}:
\begin{align}
S_{\rm np}^{\rm LY} = (g_1 + g_2 \ln \frac{Q}{2Q_0}) b_\perp^2 + g_3 \ln (100x_1x_2) |b_\perp|.
\end{align}

\end{itemize}
In the above parameterizations, $g_i$ and $Q_0$ are free parameters. For $Z^0$-boson production, $Q = M_Z$ is simply a constant and $x_1x_2 \simeq M_Z^2/s$ at low transverse momentum. Therefore, the $Q$ and $x_1x_2$ dependences in the above parameterizations can be translated into the $\sqrt{s}$-dependence. 

To summarize, both the DWS and the BLNY parameterizations adopt a pure Gaussian form, except the coefficient in the BLNY parameterization depends on $\sqrt{s}$ while that in the DWS parameterization does not. The LY parameterization adds a linear term to the pure Gaussian form while in the SIYY parameterization, a logarithmic term is included. In this paper, we  employ the following three forms of parameterization,
\begin{align}
 S_{\rm np}^{\rm Gaussian} & = g_a (Q=M_Z,\sqrt{s}) b_\perp^2,
\\
 S_{\rm np}^{\rm Gaussian+Log} & = g_a (Q=M_Z,\sqrt{s}) b_\perp^2 \nonumber \\
& + g_b (Q=M_Z,\sqrt{s}) \ln \frac{|b_\perp|}{b_*},
\\
 S_{\rm np}^{\rm Gaussian+Linear} & = g_a (Q=M_Z,\sqrt{s}) b_\perp^2 \nonumber \\
& + g_b(Q=M_Z,\sqrt{s}) |b_\perp|.
\end{align}
Here, the $Q$ and $x_1x_2$ dependences have been absorbed into $g_{a}$ and $g_b$. For instance, in the DWS parameterization, $g_a^{\rm DWS} = g_1 + g_2 \ln M_Z/2Q_0$ and in the BLNY parameterization, $g_a^{\rm BLNY} = g_1 + g_2 \ln M_Z/2Q_0 + g_3 \ln (100M_Z^2/s)$.
Although this work will not reveal information on the $Q$-dependence of these free parameters, we can study the $\sqrt{s}$-denpendence in great detail since we have eliminated the other variable.

\section{Numerical Results}

In this work, we calculate the $\phi^*$-distribution of $Z^0$-boson production in $pp$ collisions with a set of values for the free parameters and compare the results with the experimental data measured by the D0 collaboration \cite{Abazov:2010mk} at Tevatron and the ATLAS collaboration \cite{Aad:2012wfa} at the LHC to obtain the corresponding $\chi^2$ values. The absolute magnitude of the differential cross section also depends on the uncertainty in the collinear PDFs which does not interest us. In this work, we only compare the self-normalized $\phi^*$ distribution which is mainly controlled by the Sudakov factor.

\subsection{Gaussian Fit}

\begin{figure}[htb]
\includegraphics[width=0.35\textwidth]{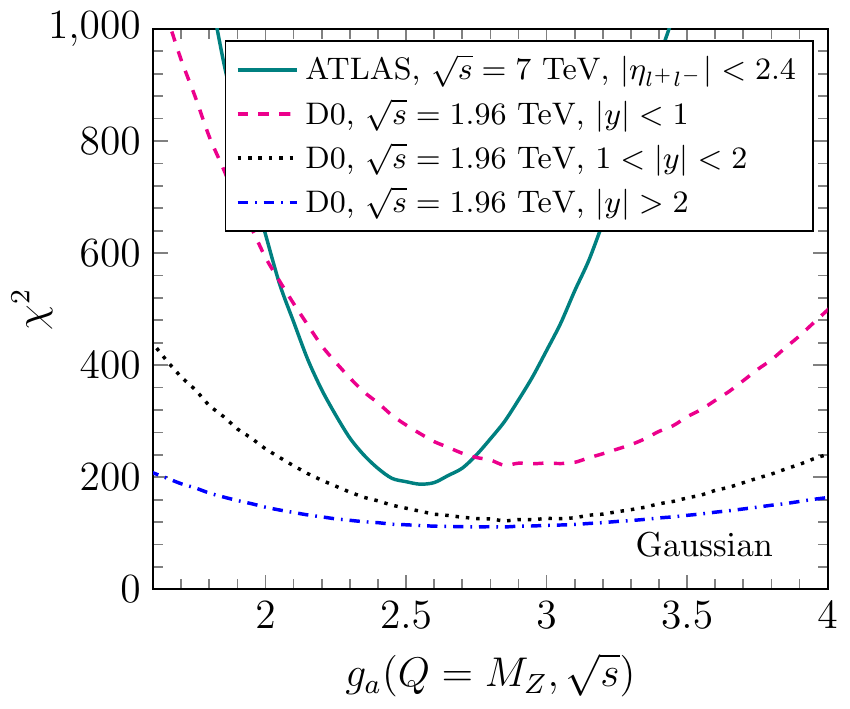}
\caption{$\chi^2$ as a function of $g_a(Q=M_Z,\sqrt{s})$ in the Gaussian fit.}
\label{fig:chi2-gaussian}
\end{figure}

We first show our results with the simple Gaussian parameterization in Fig.~\ref{fig:chi2-gaussian}. At the LHC energy \cite{Aad:2012wfa}, the value of $\chi^2$ hits the minimum at around $g_a(Q=M_Z,\sqrt{s}=7{\rm ~TeV}) \simeq 2.5$ GeV$^2$. This value is larger than that extracted with the DWS parameterization in Ref.~\cite{Landry:2002ix} which gives $g_a^{\rm DWS} (Q=M_Z) = 1.8$ GeV$^2$, but is smaller than that extracted with the BLNY parameterization~\cite{Landry:2002ix} which gives $g_a^{\rm BLNY} (Q=M_Z, \sqrt{s}=7{\rm ~TeV}) = 3.0$ GeV$^2$. 

The measurements of D0 collaboration \cite{Abazov:2010mk} are performed in three rapidity ranges. In a more forward/backward rapidity region, the statistics become smaller and therefore the $\chi^2$ curve turns into flatter. Nonetheless, all three $\chi^2$-curves show that the minimal value of $\chi^2$ resides at the vicinity of $g_a(Q=M_Z,\sqrt{s}=1.96{\rm ~TeV}) \simeq 3.0$ GeV$^2$, which is larger than the value extracted from the ATLAS measurement. This shows $g_a$ in the Gaussian fit indeed depends on $\sqrt{s}$ or, in another word, $x_1x_2$. However, in the BLNY parameterization, $g_a^{\rm BLNY} (Q=M_Z, \sqrt{s}=1.96{\rm ~TeV}) = 2.7$ GeV$^2$, which illustrates an opposite tendency. We shall not jump the gun to conclude which one is better, since only two experimental measurements have been discussed here. But this study has already demonstrated the exceptional potential of exploring non-perturbative Sudakov factor from $Z^0$-boson production. We expect more data to be released at different collision energies in the future and this issue will get clarified ultimately.

It is also interesting to note that $g_a$ in the Gaussian fit is rapidity irrelevant. This indicates that the only possible candidate of parton momentum fraction dependence of $g_a$ is $x_1x_2$ instead of $x_1^\lambda + x_2^\lambda$. A similar feature has also been observed in Ref.~\cite{Guzzi:2013aja,Su:2014wpa}.

\begin{figure}[htb]
\includegraphics[width=0.35\textwidth]{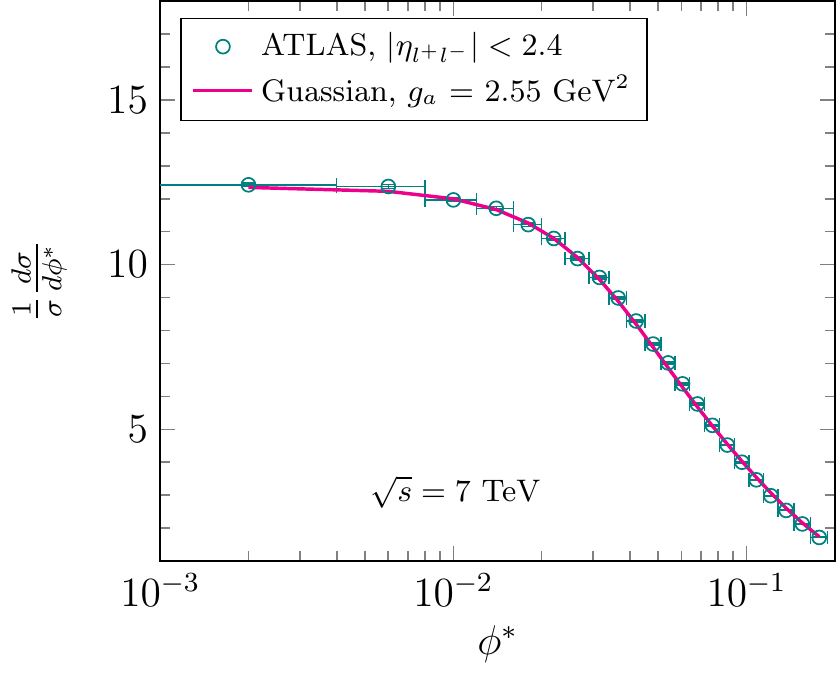}
\caption{Self-normalized $\phi^*$ distribution calculated with the Gaussian fit compared with ATLAS data \cite{Aad:2012wfa}.}
\label{fig:phistar-gaussian}
\end{figure}

The $\chi^2$ values are very large in Fig.~\ref{fig:chi2-gaussian}. This is mainly because the experimental data are extremely precise. We show our results in this Gaussian fit compared with the ATLAS data \cite{Aad:2012wfa} in Fig.~\ref{fig:phistar-gaussian}.

\subsection{Gaussian+Log Fit}

In the SIYY parameterization \cite{Su:2014wpa}, an additional logarithmic term is employed. In their extraction, $b_{\rm max} = 1.5$ GeV$^{-1}$, which is larger than the value we are currently using. Effectly, more effects have been deposited into the perturbative Sudakov factor in their study and a direct comparison to our work is inappropriate. For consistency of this paper, we will still utilize $b_{\rm max} = 0.5$ GeV$^{-1}$ in the Gaussian+Log fit.

\begin{figure}[htb]
\includegraphics[width=0.24\textwidth]{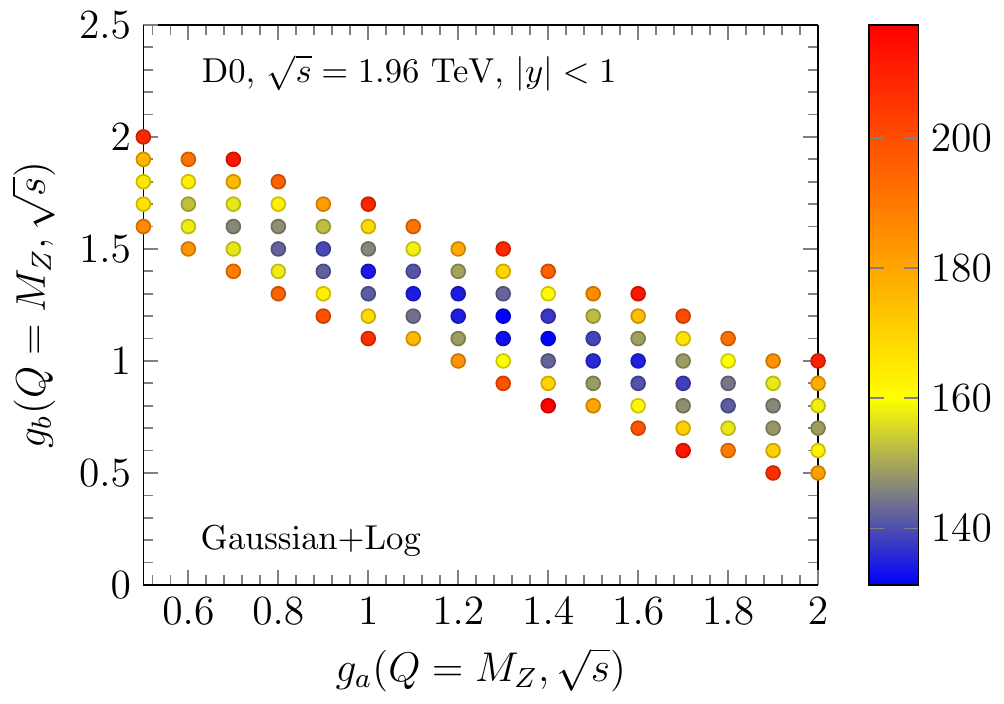}\includegraphics[width=0.24\textwidth]{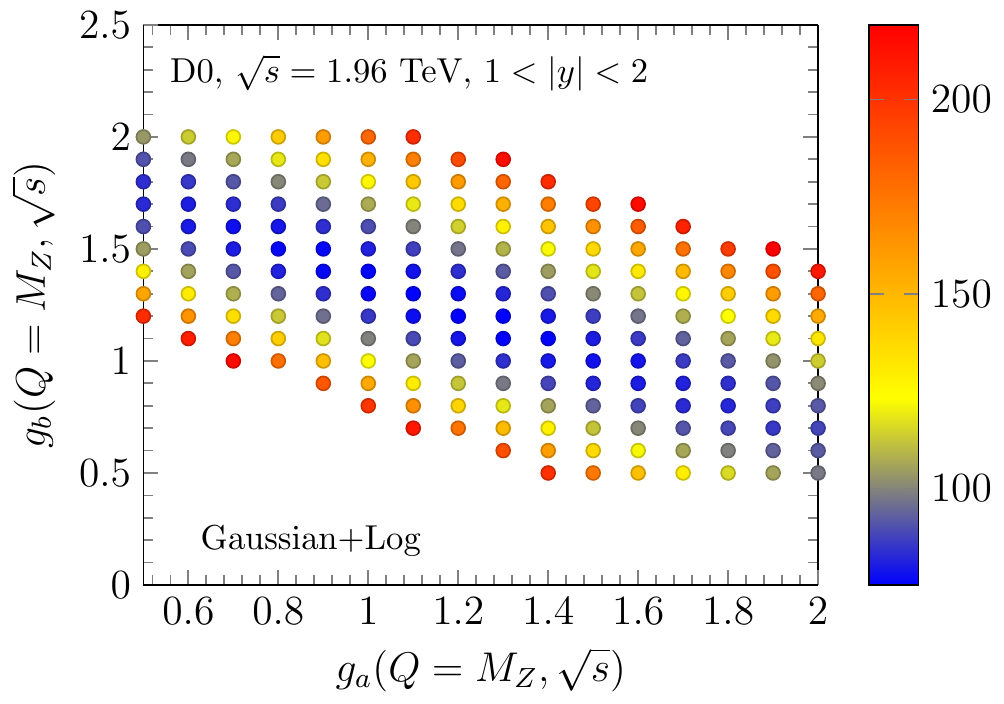} \\
\includegraphics[width=0.24\textwidth]{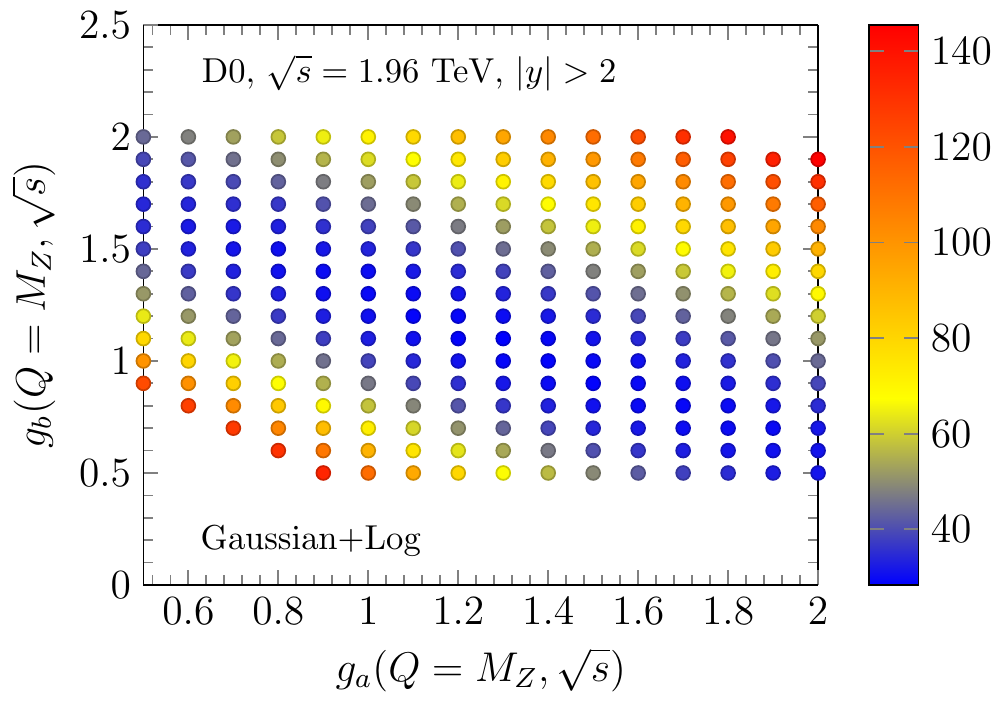}\includegraphics[width=0.24\textwidth]{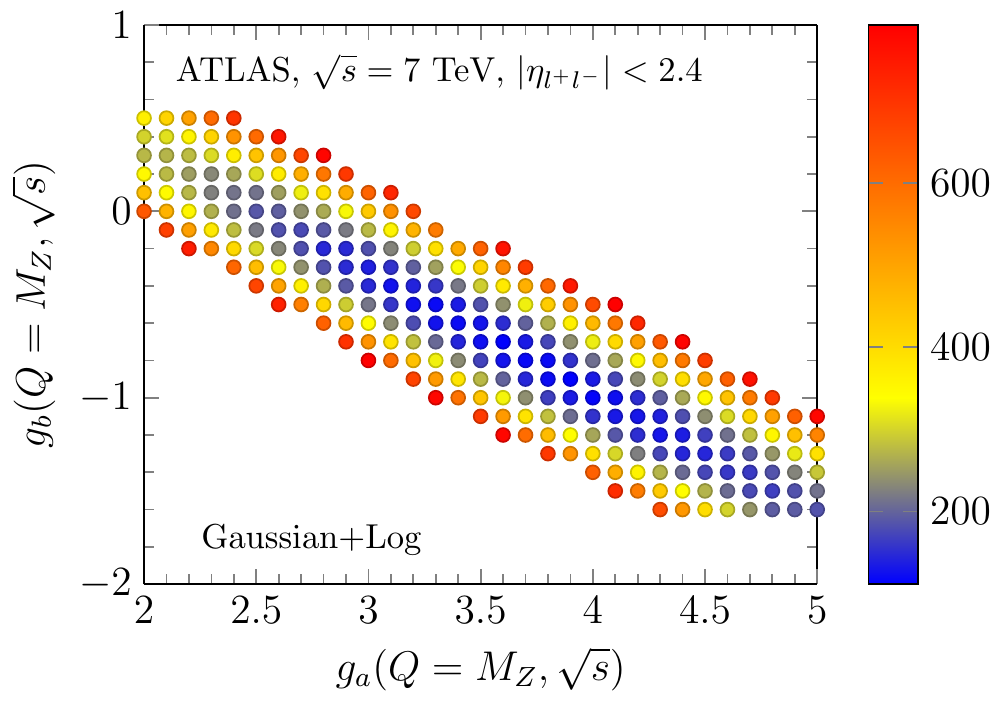}
\caption{$\chi^2$ as a function of $g_1$ and $g_2$ in the Gaussian+Log parameterization.}
\label{fig:chi2-log}
\end{figure}

In Fig.~\ref{fig:chi2-log}, we show our $\chi^2$-plots at $\sqrt{s} = 1.96$ TeV and $\sqrt{s}=7$ TeV in the Gaussian+Log parameterization. The values of $\chi^2$ are represented by different colors. The D0 data \cite{Abazov:2010mk} at different rapidities have well confined the minimal $\chi^2$ values in the same lake where $g_{a} (Q=M_Z,\sqrt{s}=1.96{\rm ~TeV}) \simeq 1.3$ GeV$^2$ and $g_{b} (Q=M_Z,\sqrt{s}=1.96{\rm ~TeV}) \simeq 1.4$. There is no rapidity dependence in these parameters. At $\sqrt{s}=7$ TeV, $g_{a}$ grows much larger and $g_{b}$ declines to negative. We obtain $g_{a} (Q=M_Z,\sqrt{s}=7{\rm ~TeV}) \simeq 3.9$ GeV$^2$ and $g_{b} (Q=M_Z,\sqrt{s}=1.96{\rm ~TeV}) \simeq -0.9$. The difference between the best fits in these two experiments reveals the $\sqrt{s}$-dependence, or $x_1x_2$-dependence equivalently speaking, of these two free parameters.
 
\begin{figure}[htb]
\includegraphics[width=0.24\textwidth]{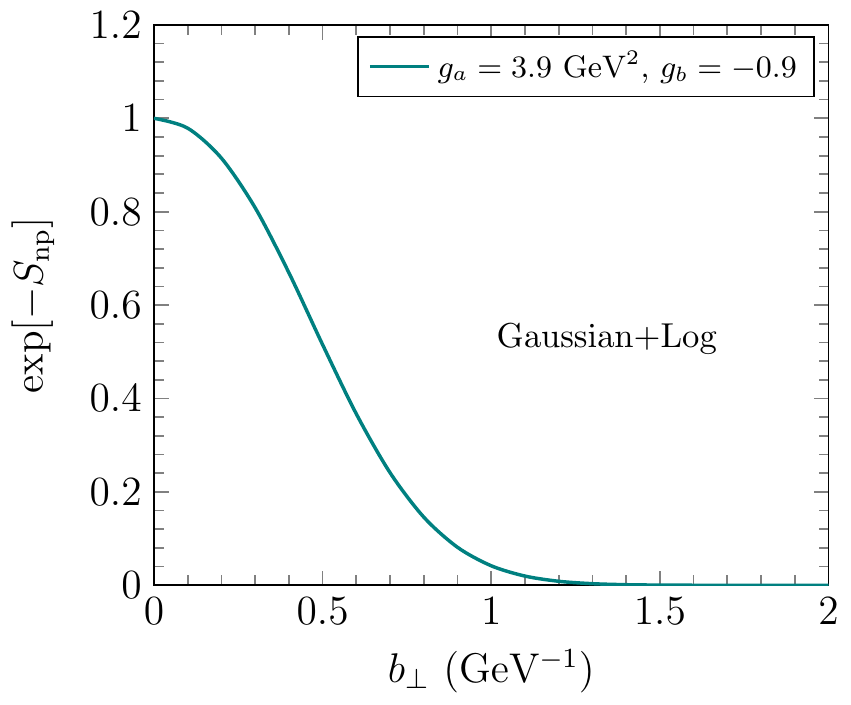}\includegraphics[width=0.24\textwidth]{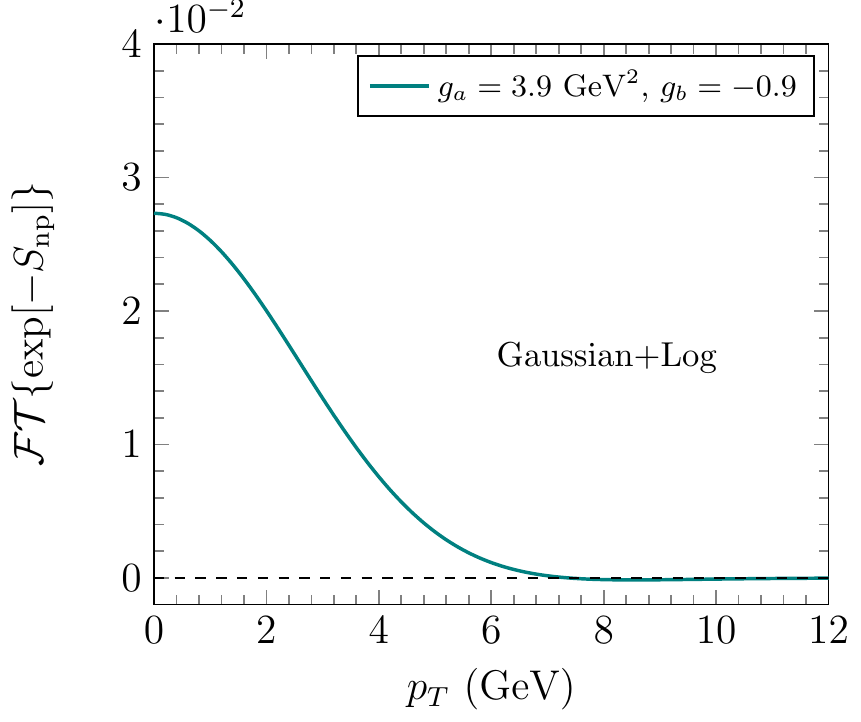}
\caption{Non-perturbative Sudakov factor in the Gaussian+Log fit and its Fourier transform.}
\label{fig:snp-log}
\end{figure}

In the language of $k_t$-dependent factorization, the Fourier transform of the non-perturbative Sudakov factor can be interpreted as the quark transverse momentum distribution at the initial scale. Although the negative logarithmic term is quite small, it still breaks the positivity of distribution at large transverse momentum. In Fig.~\ref{fig:snp-log}, we show the non-perturbative Sudakov factor in the Gaussian+Log fit and its Fourier transform which becomes negative at $p_T \ge 7.4$ GeV. However the violation only occurs at the initial scale and the magnitude is so small that it is barely visiable with naked eye. The contribution to the final results from the non-physical region should be negligible. Therefore, the parameters extracted here can still be employed in the future studies.

\begin{figure}[htb]
\includegraphics[width=0.35\textwidth]{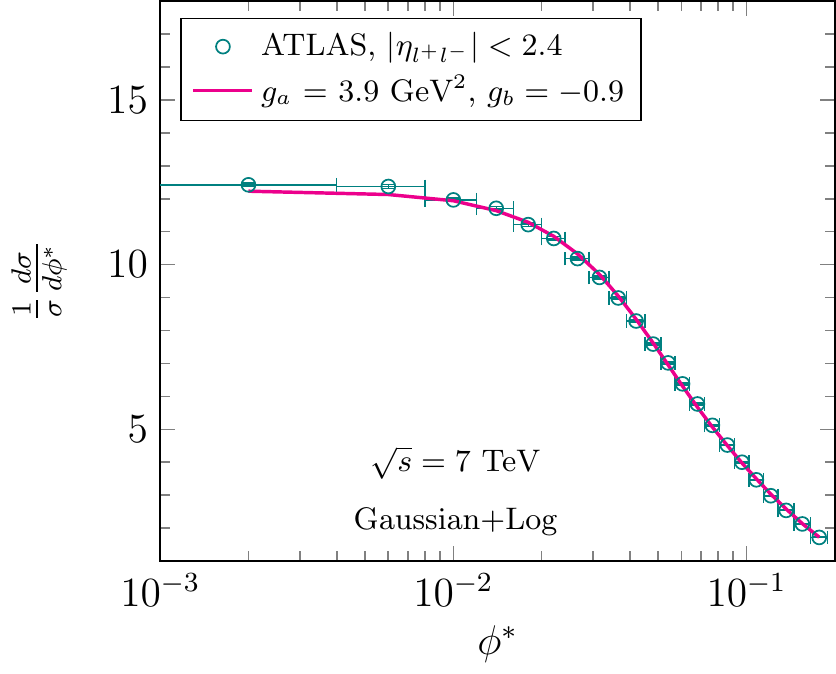}
\caption{Self-normalized $\phi^*$ distribution calculated with the Gaussian+Log fit compared with ATLAS data \cite{Aad:2012wfa}.}
\label{fig:phistar-log}
\end{figure}

In Fig.~\ref{fig:phistar-log}, we show our results in the Gaussian+Log fit compared with the ATLAS data \cite{Aad:2012wfa}. If the strict positivity constraint is enforced, $g_b$ cannot be negative and then we find the minimal $\chi^2$ locates at $g_a(Q=M_Z,\sqrt{s}=7{\rm ~TeV})=2.5$ GeV$^2$ and $g_b(Q=M_Z,\sqrt{s}=7{\rm ~TeV})=0$ which is the same with that in the pure Gaussian fit.

\subsection{Gaussian+Linear Fit}

\begin{figure}[htb]
\includegraphics[width=0.24\textwidth]{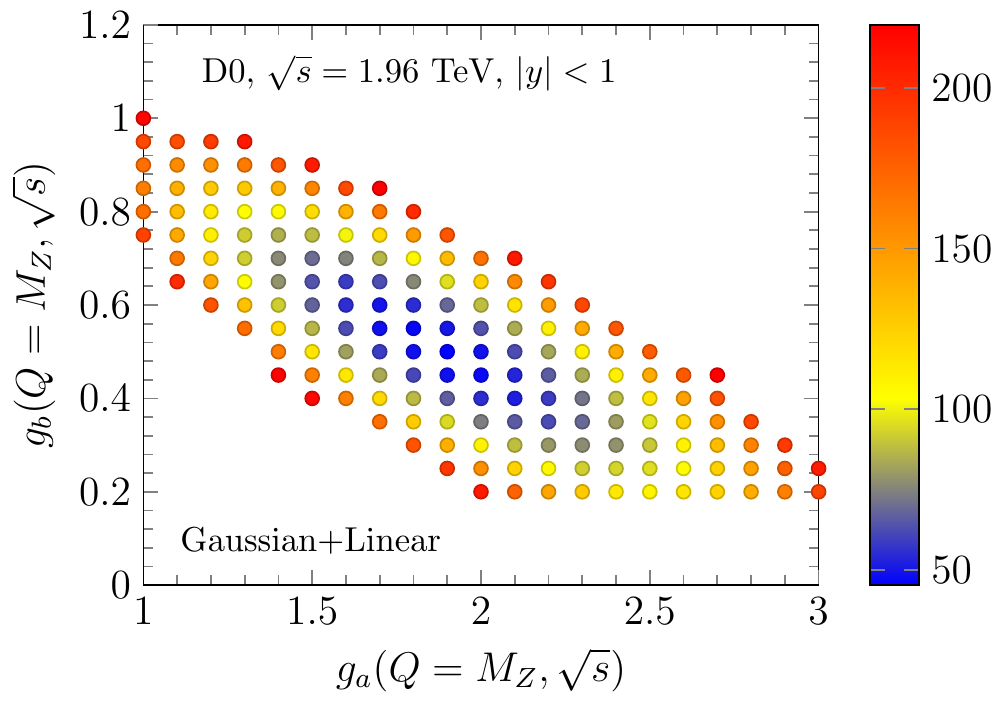}\includegraphics[width=0.24\textwidth]{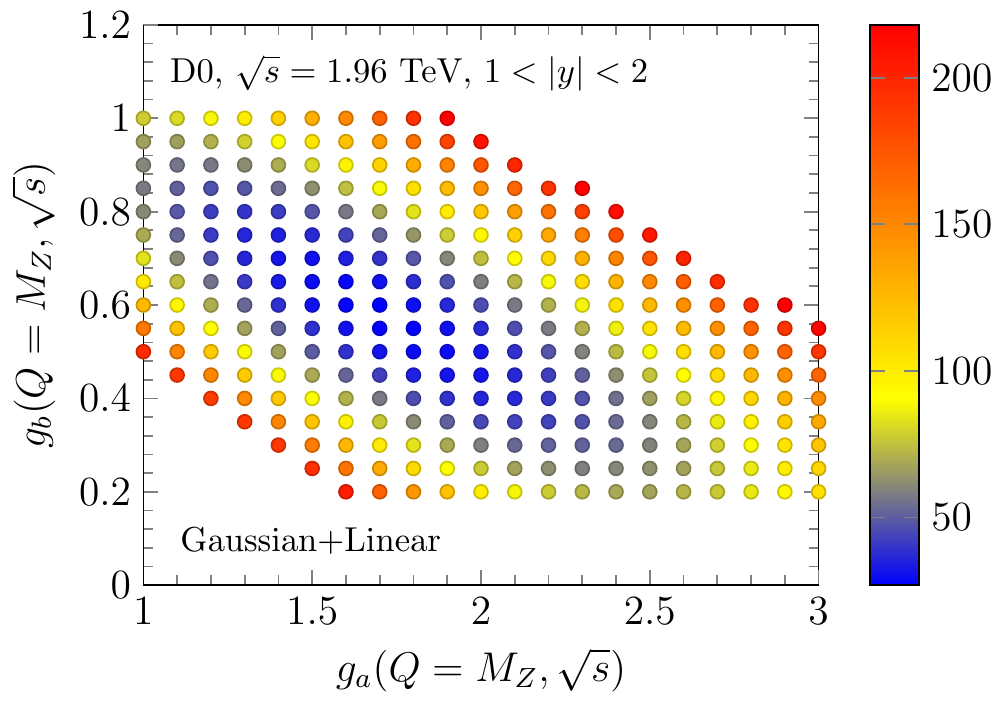}
\\
\includegraphics[width=0.24\textwidth]{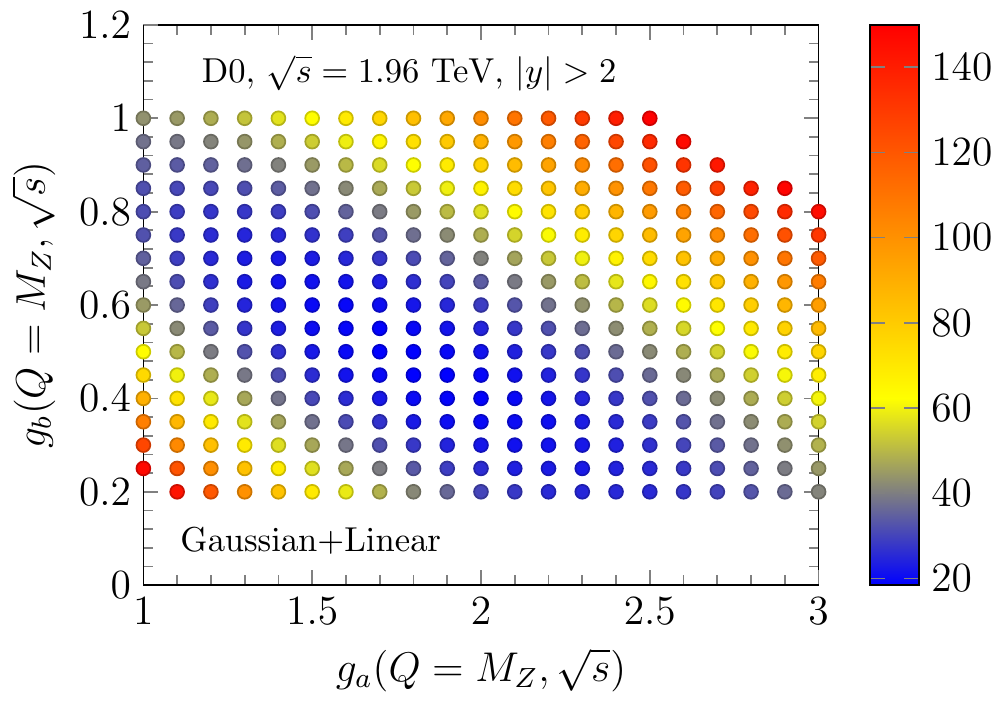}\includegraphics[width=0.24\textwidth]{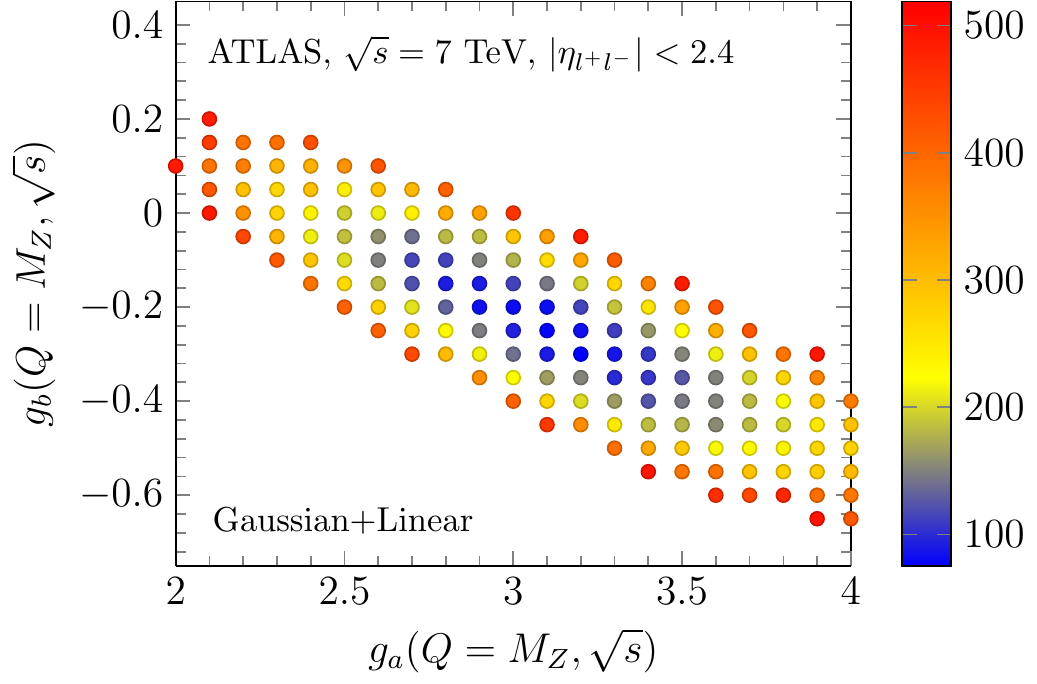}
\caption{$\chi^2$ as a function of $g_1$ and $g_2$ in the Gaussian+Linear parameterization.}
\label{fig:chi2-linear}
\end{figure}

Finally, we show the 3D $\chi^2$ plots calculated with the Gaussian+Linear fit in Fig.~\ref{fig:chi2-linear} at $\sqrt{s}=1.96$ TeV and $\sqrt{s}=7$ TeV. The $\chi^2$ value finds its minimum at the pond where $g_a(Q=M_Z,\sqrt{s}=1.96{\rm ~TeV}) \simeq 1.8$ GeV$^2$ and $g_b(Q=M_Z,\sqrt{s}=1.96{\rm ~TeV}) \simeq 0.5$ GeV at all three rapidities of the D0 measurements. The values extracted in this work are slightly different with those in Ref.~\cite{Ladinsky:1993zn}, which gives $g_a^{\rm LY}(Q=M_Z,\sqrt{s}=1.96{\rm ~TeV})=2.1$ GeV$^2$ and $g_b^{\rm LY}(Q=M_Z, \sqrt{s}=1.96{\rm ~TeV})=0.25$ GeV, and also are different with those in a more updated work \cite{Landry:2002ix}, which gives $g_a^{\rm LY}(Q=M_Z,\sqrt{s}=1.96{\rm ~TeV})=1.9$ GeV$^2$ and $g_b^{\rm LY}(Q=M_Z, \sqrt{s}=1.96{\rm ~TeV})=0.046$ GeV.

At $\sqrt{s}=7$ TeV, a bulky $\chi^2$-analysis shows that $g_a$ becomes much larger and $g_b$ turns into negative. We obtain $g_a(Q=M_Z,\sqrt{s}=7{\rm ~TeV}) \simeq 3.2$ GeV$^2$ and $g_b(Q=M_Z,\sqrt{s}=7{\rm ~TeV}) \simeq -0.3$ GeV. However, if the non-perturbative Sudakov factor is interpreted as the intrinsic transverse momentum distribution of quark, this negative linear term slightly violates the positivity of the distribution at large transverse momentum. In light of the strict positivity constraint, the minimal $\chi^2$ value locates at $g_b=0$ GeV. This Gaussian+Linear parameterization becomes the de facto Gaussian form at $\sqrt{s}=7$ TeV again.

\begin{figure}[htb]
\includegraphics[width=0.24\textwidth]{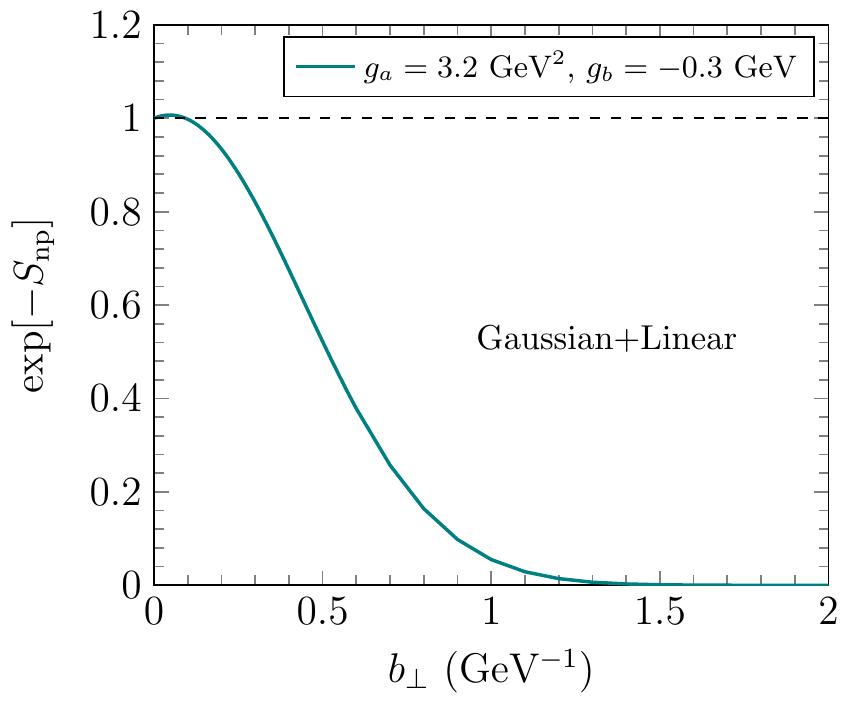}\includegraphics[width=0.24\textwidth]{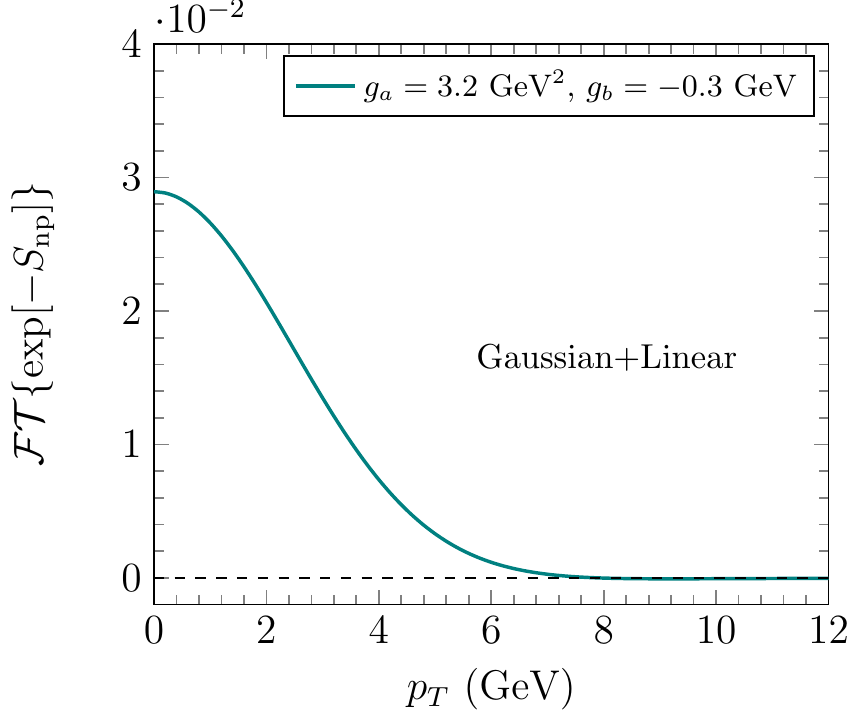}
\caption{Non-perturbative Sudakov factor in the Gaussian+Linear fit and its Fourier transform.}
\label{fig:snp-linear}
\end{figure}

\begin{figure}[h!]
\includegraphics[width=0.35\textwidth]{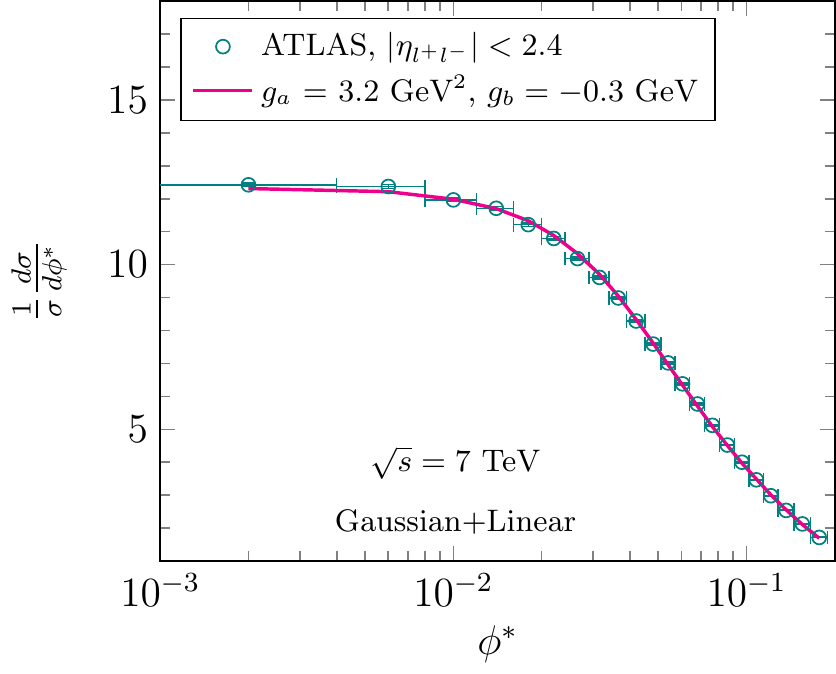}
\caption{Self-normalized $\phi^*$ distribution calculated with the Gaussian+Linear fit compared with ATLAS data \cite{Aad:2012wfa}.}
\label{fig:phistar-linear}
\end{figure}

Similar to the case in the Gaussian+Log fit, the magnitude of this violation is very small. In Fig.~\ref{fig:snp-linear}, we show the non-perturbative Sudakov factor in the Gaussian+Linear fit and its Fourier transform. The Fourier transform becomes negative with a tiny absolute magnitude at $p_T \ge 8$ GeV. This non-physical region has little impact on the final results. 
In Fig.~\ref{fig:phistar-linear}, we show our results in the Gaussian+Linear fit compared with the ATLAS data \cite{Aad:2012wfa}.

\section{Summary}

Although the differential cross section of $Z^0$-boson production is not very sensitive to non-perturbative physics, the ultra precise $\phi^*$-distribution still provides an indispensable opportunity to explore the non-perturbative Sudakov factor. 

We have calculated the perturbative Sudakov factor up to the NLL accuracy with $b_{\rm max}=0.5$ GeV$^{-1}$. For the non-perturbative Sudakov factor, we have employed the Gaussian, Gaussian+Log and Gaussian+Linear parameterizations and extract the corresponding free parameters with a $\chi^2$-analysis. In general we find the values of the $\chi^2$ in the pure Gaussian fit are the largest since there is only one degree of freedom. Adding a linear $b_\perp$-dependent term to the pure Gaussian fit can significantly reduce the value of $\chi^2$ which shows this Gaussian+Linear fit is the best form of parameterization so far. The $\chi^2$ value can be further reduced as long as more degrees of freedom are introduced. This data-driven analysis is another topic that can be proceeded once more data are available.

In the Gaussian fit, we find $g_a \simeq 3.0$ GeV$^2$ at $\sqrt{s}=1.96$ TeV and it becomes smaller at $\sqrt{s}=7$ TeV which gives  $g_a \simeq 2.5$ GeV$^2$. In both Gaussian+Log and Gaussian+Linear fits, $g_a$ becomes larger and $g_b$ becomes smaller at a larger $\sqrt{s}$. In the Gaussian+Log fit, $g_a \simeq 1.3$ GeV$^2$ and $g_b \simeq 1.4$ at $\sqrt{s}=1.96$ TeV and $g_a \simeq 3.9$ GeV$^2$ and $g_b \simeq -0.9$ at $\sqrt{s} = 7$ TeV. In the Gaussian+Linear fit, $g_a \simeq 1.8$ GeV$^2$ and $g_b \simeq 0.5$ GeV at $\sqrt{s}=1.96$ TeV and $g_a \simeq 3.2$ GeV$^2$ and $g_b \simeq -0.3$ GeV at $\sqrt{s} = 7$ TeV. 

All these parameters have no rapidity-dependence, which indicates that the parton momentum fraction dependence favors the $x_1x_2$ form to $x_1^\lambda + x_2^\lambda$. This observation is consistent with that in a previous study in Ref.~\cite{Su:2014wpa}.

\begin{acknowledgments}

We would like to thank C. Marquet for drawing our attention to the $\phi^*$ distribution. Part of the numerical evaluation in this paper was accomplished on the computer cluster of CPHT, Ecole Polytechnique. We would like to express our gratitude to the IT department of CPHT for extending our permit to access their cluster.

\end{acknowledgments}

\end{document}